\documentclass[conference]{IEEEtran}
\IEEEoverridecommandlockouts

\usepackage{algorithmic}
\usepackage{amsmath,amssymb,amsfonts}
\usepackage{array}
\usepackage{bm}
\usepackage{cite}
\usepackage[T1]{fontenc}
\usepackage{gensymb}
\usepackage{graphicx}
\usepackage{epstopdf}
\usepackage[utf8]{inputenc}
\usepackage{pdfpages}
\usepackage{subcaption}
\usepackage{textcomp}
\usepackage{xcolor}

\DeclareCaptionLabelFormat{andtable}{#1~#2  \&  \tablename~\thetable}

\def\BibTeX{{\rm B\kern-.05em{\sc i\kern-.025em b}\kern-.08em
    T\kern-.1667em\lower.7ex\hbox{E}\kern-.125emX}}

\graphicspath{{./images/}}

\begin{document}

\title{Channel Extrapolation for FDD Massive MIMO: Procedure and Experimental Results}
\author{
    \IEEEauthorblockN{Thomas~Choi\IEEEauthorrefmark{1}, Fran\c{c}ois~Rottenberg\IEEEauthorrefmark{1}, Jorge Gomez-Ponce\IEEEauthorrefmark{1}, Akshay Ramesh\IEEEauthorrefmark{1}, Peng Luo\IEEEauthorrefmark{1}}
    \IEEEauthorblockN{Jianzhong Zhang\IEEEauthorrefmark{2}, and Andreas F. Molisch\IEEEauthorrefmark{1}}
    \IEEEauthorblockA{\IEEEauthorrefmark{1}University of Southern California, Los Angeles, CA, USA}
    \IEEEauthorblockA{\IEEEauthorrefmark{2}Samsung Research America, Richardson, TX, USA}
}
\maketitle

\begin{abstract}
Application of massive multiple-input multiple-output (MIMO) systems to frequency division duplex (FDD) is challenging mainly due to the considerable overhead required for downlink training and feedback. Channel extrapolation, \textit{i.e.}, estimating the channel response at the downlink frequency band based on measurements in the disjoint uplink band, is a promising solution to overcome this bottleneck. This paper presents measurement campaigns obtained by using a wideband (350 MHz) channel sounder at 3.5 GHz composed of a calibrated 64 element antenna array, in both an anechoic chamber and outdoor environment. The Space Alternating Generalized Expectation-Maximization (SAGE) algorithm was used to extract the parameters (amplitude, delay, and angular information) of the multipath components from the attained channel data within the ``training'' (uplink) band. The channel in the downlink band is then reconstructed based on these path parameters. The performance of the extrapolated channel is evaluated in terms of mean squared error (MSE) and reduction of beamforming gain (RBG) in comparison to the ``ground truth'', \textit{i.e.}, the {\em measured} channel at the downlink frequency. We find strong sensitivity to calibration errors and model mismatch, and also find that performance depends on propagation conditions: LOS performs significantly better than NLOS.
\end{abstract}

\section{Introduction}
Massive multiple-input multiple-output (MIMO) will be an essential part of 5G and beyond wireless communications systems. While massive MIMO performs generally better using time division duplex (TDD) than frequency division duplex (FDD)\cite{TDDvsFDD}, frequency regulators over the world have assigned large swathes of valuable spectrum as band {\em pairs} for FDD operation; furthermore backward compatibility considerations (\textit{e.g.}, with Long-Term Evolution (LTE)) also often require FDD operation. Enabling FDD massive MIMO has thus been a popular research topic in recent years \cite{FDD_Zhang, FDD_Love, FDD_Molisch}. The key challenge lies in the acquisition of channel state information for the downlink. Current systems use downlink training and feedback from the user equipment (UE) to the base station (BS), which leads to considerable overhead. An alternative is frequency extrapolation, \textit{i.e.}, using channel information measured in the uplink frequency band to estimate  channel information in the downlink frequency band.  

The idea of frequency extrapolation for MIMO systems has been studied in the past. One avenue of research was the extrapolation of {\em second-order channel characteristics} such as angular power spectra and correlation matrices. It can be argued from physical considerations that the directions from which the dominant powers are coming change only weakly with frequency. Angular power spectra derived from Bartlett and Capon beamformers were used in \cite{Salous_FDD_MIMO} for $4 \times 4$ MIMO. The authors in \cite{Rice_FDD} performed frequency extrapolation of extracted dominant angles using maximum-likelihood directional estimators for 64-element arrays and combined them with an ingenious training/feedback scheme.  

Another, more difficult, goal is the extrapolation of the {\em complex, instantaneous} channel frequency response. Since the phases of the multipath components (MPCs) change rapidly over frequency, such an extrapolation is extremely sensitive to both measurement errors within the training band, and errors in the extrapolation algorithm. While theoretical investigations and experiments based on synthetic channel models have been performed for a while \cite{synthetic1, synthetic2, synthetic3, rottenberg2019channel}, evaluations based on indoor and outdoor measurements have been done only recently. 
In particular, extrapolation based on a Fourier representation of the channel using uniform linear arrays for 4 to 16 antenna elements was done in \cite{MIT_FDD}. Deep learning based frequency extrapolation with 32 antenna elements showed sum-rates information for measured MIMO cases in \cite{deeplearning_FDD}. In all these cases, the ``training bandwidth'' (\textit{i.e.}, the bandwidth in which the channel was measured for estimation), was between 10 to 20 MHz, while the extrapolation range was between 40 to 72 MHz. Most of these investigations present the results in terms of achievable beamforming gain, SNR, or data rate.   

In this paper, we present results from a measurement campaign in both an anechoic chamber and a realistic outdoor scenarios. Results indicate error sources such as calibration errors lead to worse performance than theoretical bounds \cite{rottenberg2019channel, Swindle}. As our investigation in \cite{rottenberg2019channel}, our extrapolation approach is based on (i) measuring with a calibrated array, (ii) extracting the MPC parameters with a high resolution parameter estimation (HRPE) algorithm (the Space Alternating Generalized Expectation-Maximization (SAGE) algorithm \cite{SAGE}), and (iii) synthesizing the channel response at the new band. 

Our measurements have the following characteristics. First, channel measurements are done with a wideband (350 MHz), real-time time-domain channel sounder that allows to mimic operation of a 5G system. Second, we perform our measurements in an outdoor environment under a variety of setups, including line-of-sight (LOS), non-LOS (NLOS), and partial LOS (PLOS), encountering real channel conditions that might deviate from the model (finite sum of plane waves) that forms the basis of the extrapolation. Third, calibration and measurements are done with different mountings (rotating positioner and measurement cart respectively), with some time between them, providing information about sensitivity to calibration ``aging'' and its impact on the extrapolation. Last but not least, we investigate different metrics to judge the quality of the extrapolation, especially comparing mean squared error (MSE) and reduction of beamforming gain (RBG).    

The paper is organized as follows. Section \ref{cal_meas} describes channel sounder specifications, antenna array calibration, and measurement scenarios. Section \ref{MSE} describes the process of applying SAGE to the measured data and extrapolating the channel. Measurement results in Section \ref{Results} quantify the resulting MSE and RBG of estimated channels in the extrapolated frequency bands, followed by their implications for practical channel extrapolation in the frequency domain. Section \ref{Conclusion} provides the conclusions and indications of our future work.

\section{Sounder, Calibration, and Measurement} \label{cal_meas}

\subsection{Channel Sounder Specifications}
\begin{figure}[h]
    \centering
    \includegraphics[scale = 0.068]{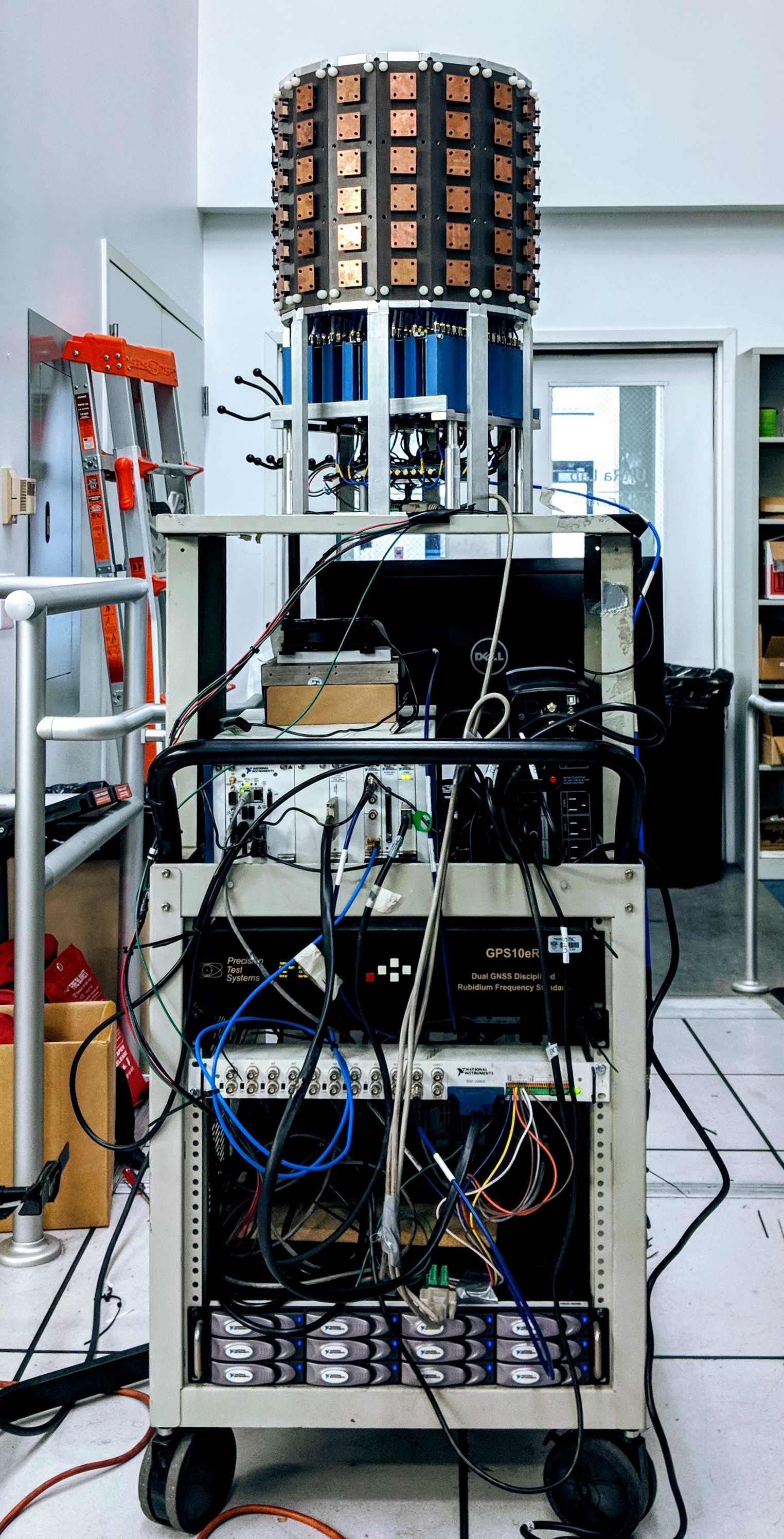}
    \begin{tabular}[b]{  m{2.05cm} | m{0.95cm}  }
    \hline
    \multicolumn{2}{c}{Hardware Specs} \\
    \hline
    \hline
    \scriptsize{Center frequency} & \scriptsize{3.5 GHz}\\
    \hline
    \scriptsize{\# of TX/RX ports} & \scriptsize{1/64}\\
    \hline
    \scriptsize{TX EIRP} & \scriptsize{30 dBm}\\
    \hline
    \scriptsize{RX beamwidth az/el} & \scriptsize{50\degree/100\degree}\\
    \hline
    \scriptsize{RX RF switch time} & \scriptsize{100 ns}\\
    \hline
    \scriptsize{RX total noise figure} & \scriptsize{$<$ 11 dB}\\
    \hline
    \multicolumn{2}{c}{Sounding Waveform} \\
    \hline
    \scriptsize{Bandwidth} & \scriptsize{350 MHz}\\
    \hline
    \scriptsize{Sample frequency} & \scriptsize{1.25GSps}\\
    \hline
    \scriptsize{\# of subcarriers} & \scriptsize{2801}\\
    \hline
    \scriptsize{PAPR} & \scriptsize{0.4 dB}\\
    \hline
    \scriptsize{Waveform duration} & \scriptsize{8 us}\\
    \hline
    \scriptsize{Waveform repetition} &    \scriptsize{10}\\
    \hline
    \scriptsize{SISO duration} & \scriptsize{80 us}\\
    \hline
    \scriptsize{SIMO duration} & \scriptsize{10.24 ms}\\
    \hline
  \end{tabular}
  \captionlistentry[table]{A table beside a figure}
    \captionsetup{labelformat=andtable}
    \caption{USC SIMO channel sounder}
    \label{Sounder}
\end{figure}

Massive MIMO usually refers to a BS with a large number of antenna elements communicating with a multitude of single-antenna UEs. Since the channel estimation errors for the different UEs are independent of each other provided that they use orthogonal training sequences, it is sufficient to analyze a single-input multiple-output (SIMO) setup; such a setup was used for our real-time channel sounder. Figure \ref{Sounder} and Table \ref{Sounder} show the photo and the specifications of the University of Southern California (USC) SIMO channel sounder. We chose the 3.5 GHz frequency band for the sounder because it will be used widely for 5G and beyond. USC obtained an experimental license to operate at the frequency from the Federal Communications Commission (FCC).  

On the transmitter (TX) side, the sounder uses a single omni-directional antenna. The antenna has a non-uniform beampattern in elevation, with a beamwidth (full-width half-maximum, FWHM) of $\sim90 \degree$. Note that due to the SIMO setup, extraction of angles of departure is not possible, which may be one of the error sources in the extraction of the MPC parameters \cite{Landmann} when the antenna characteristics are not separable in angle and frequency; the campaign in this paper kept the error small by measuring in outdoor scenarios with limited elevation spread. 

The receiver (RX) uses a cylindrical $64$ element array. Patch antenna elements are put together in $16$ columns, where each column contains 4 active patches in the middle (forming 4 rows), plus a dummy patch (terminated with 50 ohm loads) each at the top and bottom. Each element has one vertically polarized and one horizontally polarized port. While the measurements were conducted with all 128 ports, only 64 vertically polarized ports were used for evaluation of the channel. The reason is because cross-polarized ports in RX had failed to provide consistent radiation patterns during calibration due to high noise sensitivity (TX is vertically polarized). Note that the lack of dual-polarization at TX and RX might also contribute to modeling errors \cite{Landmann}. Each vertically polarized port has azimuth beamwidth of $50\degree$ and elevation beamwidth of $100\degree$ (see Fig. \ref{fig:patterns}). The ``stacked'' patch design provides a wide bandwidth ($S_{11}< -10$ dB), $\sim10\%$ of the center frequency ($\sim$350 MHz of 3.5 GHz). 

The sounder operates according to the switched sounding principle, where a single Radio Frequency (RF) chain is sequentially connected to each of the 128 ports by a fast (100 ns switching time) electronic switch. The noise figure of the RX RF chain varies, with up to 11 dB when the variable attenuator (required for adjustments to large input level variations) in the chain is at 30 dB, and down to 2.5 dB when variable attenuator is at 0 dB.  

At the TX, an arbitrary waveform generator (AWG) creates a $350$ MHz bandwidth multitone signal containing $2801$ subcarriers (frequency spacing of $0.125$ MHz). Using waveform from \cite{PAPR}, the low peak-to-average power ratio (PAPR) of $0.4$ dB allows the system to transmit with power close to the 1 dB compression point of the power amplifier without saturation; the emitted equivalent isotropically radiated power (EIRP) is $30$ dBm. Each waveform lasts $8 \mu$s. This waveform is repeated $10$ times to increase the signal to noise ratio (SNR) by averaging, increasing the duration of one SISO measurement (\textit{i.e.}, between the TX antenna and one RX port) to $80 \mu$s. The total SIMO sounding duration is $ 128 \times 0.08 = 10.24$ ms. 

\subsection{Calibration of Antennas and RF Chains}\label{cali}

The waveform measured at the analog to digital converter (ADC) at the RX contains the transmitted waveform affected by antennas, RF chains, and the channel. Therefore, calibrations of the antennas and the RF chains are necessary in order obtain the true physical characteristics (amplitude, delay, and angular information) of the channel itself. This subsection discusses the calibration procedures; usage of the calibrated data for HRPE will be discussed in Sec. III.

The TX antenna and the RX array (including the switches) are calibrated ``together''. Two ports on the vector network analyzer (VNA) are connected to the input port of the TX antenna and output port of the switch at RX. This setup is placed in the shielded anechoic chamber at USC. TX and RX were placed at least 5 m from each other, which is larger than the Rayleigh distance of the array at the highest considered frequency ($2D^2/\lambda =$ 3 m), so that the far-field assumption holds. While TX antenna is fixed in one position, the RX array is on a rotating positioner that moves the array in $ 5\degree $ steps in both azimuth ($360\degree$) and elevation ($180\degree$), providing $72\times37$ positions. The positioner was covered with absorbers to minimize its impact on the measured array patterns. 

For each position, the array switches through the 128 ports, and for each switch position, the VNA sweeps to get frequency response per port per position. During the calibration, the TX power was 10 dBm, and a low-noise amplifier (whose impact on the frequency characteristics was eliminated in post-processing) was placed in front of the VNA receiving port, in order to increase the SNR during calibration to 50 dB. 

\begin{figure}[hbt!]
    \centering
    \begin{subfigure}[b]{0.45\textwidth}
        \centering
        \includegraphics[width=8cm]{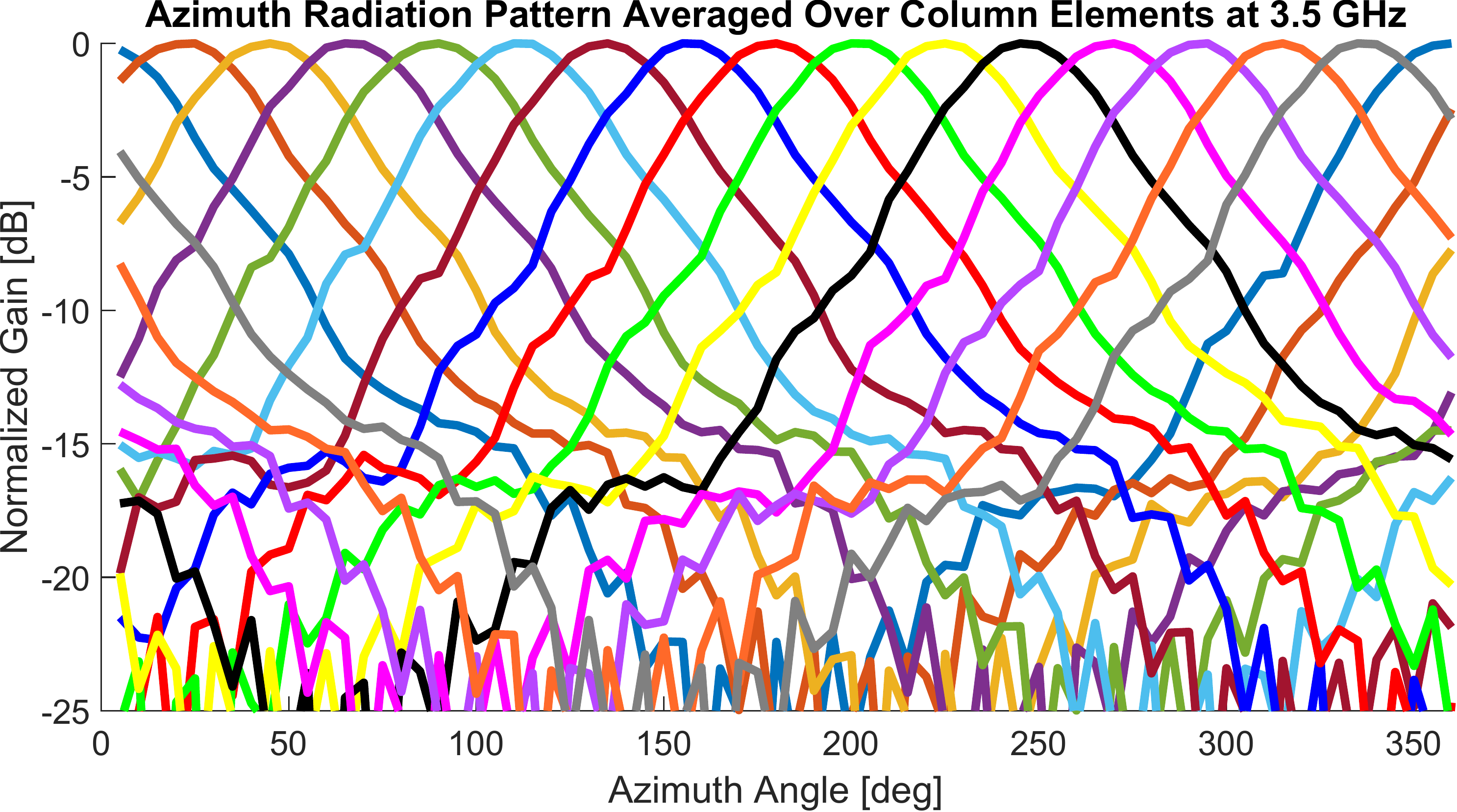}
        \caption{Azimuth}
        \vspace*{2mm}
    \end{subfigure}
    \begin{subfigure}[b]{0.45\textwidth}
        \centering
        \includegraphics[width=8cm]{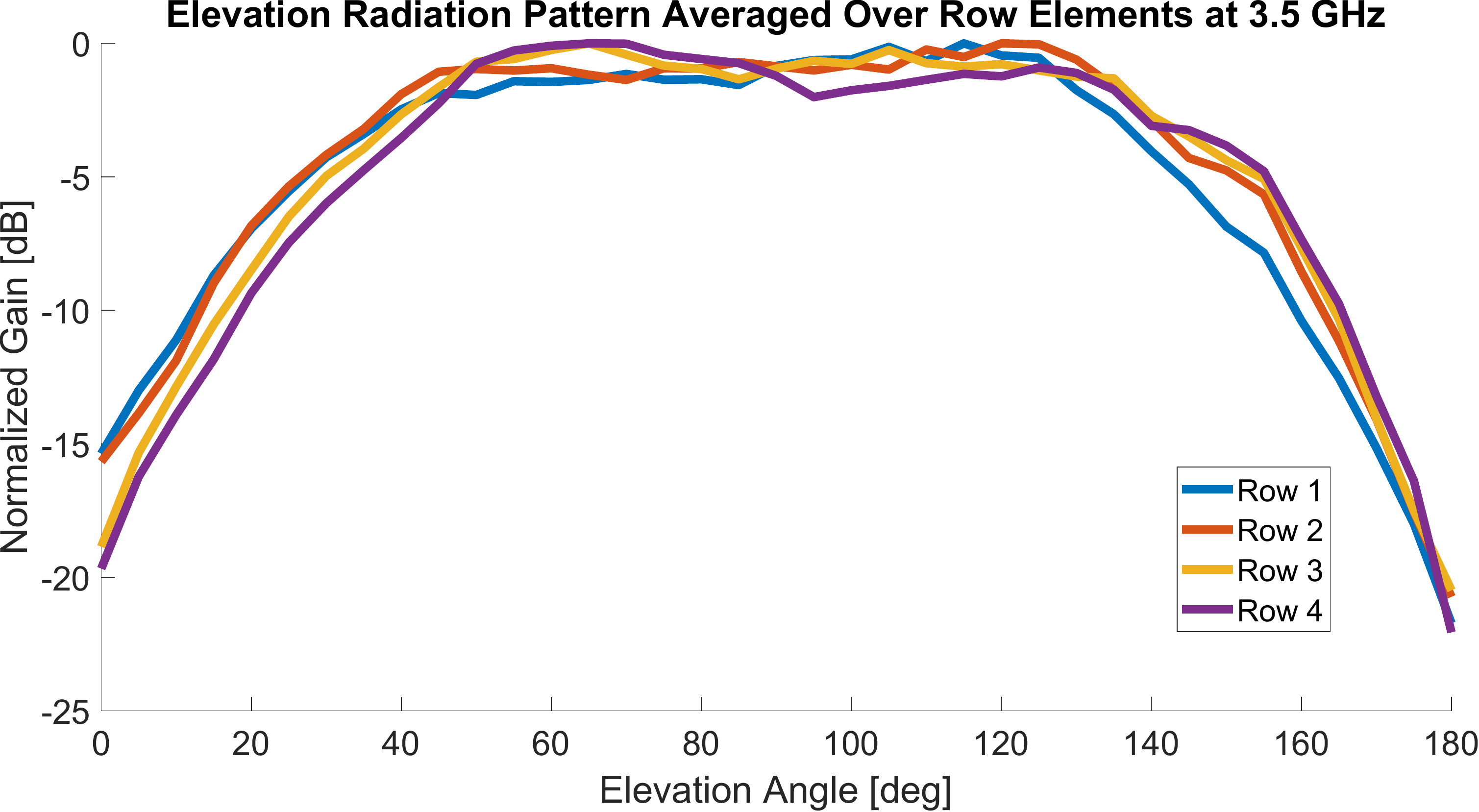} %
        \caption{Elevation}
    \end{subfigure}
\caption{Calibrated radiation patterns for a 64 V-pol ports}%
\label{fig:patterns}%
\vspace{0mm}
\end{figure}

The calibration provides a 4-Dimensional calibration matrix ($port \times azimuth \times elevation \times frequency$) characterizing the antennas. This complex  pattern $a(m,\phi, \theta, f)$, compensated by free space path loss, serves as a reference radiation pattern that is an input for the SAGE algorithm. Fig. \ref{fig:patterns} shows the averaged azimuth pattern and elevation pattern over column and row respectively at 3.5 GHz. Because frequency points measured during the calibration are fewer than the frequency points used in channel sounder (every 1 MHz vs every 0.125 MHz), the effective aperture distribution function (EADF) was used to interpolate the pattern in elevation and azimuth angles on unmeasured frequency points. \cite{Richter}.

Calibrating RF chain to obtain the so-called ``RF back-to-back'' frequency response, $H_{b2b}$, is rather simple. The TX output (at the TX antenna connector) is connected directly (via a cable) to the RX input (at the point where normally the switch output would be connected), with attenuators in between to prevent the high power output from the TX saturating the RF components in the RX. Frequency responses of the attenuators and cables are then measured separately, and compensated from the frequency response of the back-to-back measurements that include them. $H_{b2b}(f)$ is one-dimensional complex valued vector, depending only on frequency.

\begin{figure}[hbt!]
    \centering
    \begin{subfigure}[b]{0.175\textwidth}
        \centering
        \includegraphics[height=4.1cm]{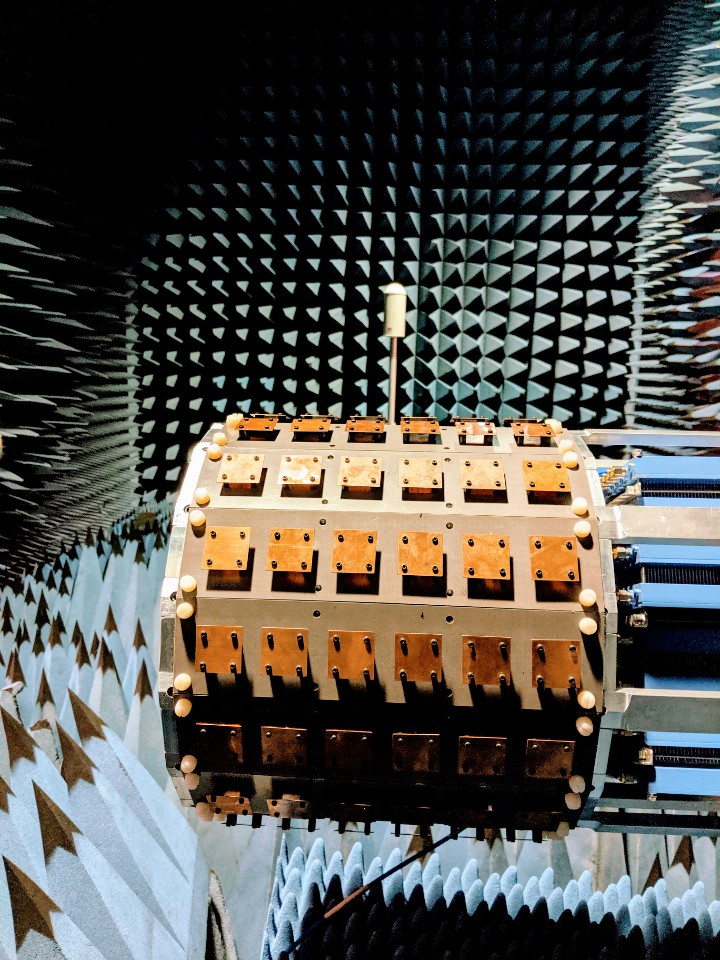}
        \caption{Chamber}
    \end{subfigure}%
    \begin{subfigure}[b]{0.27\textwidth}
        \centering        \includegraphics[height=4.1cm]{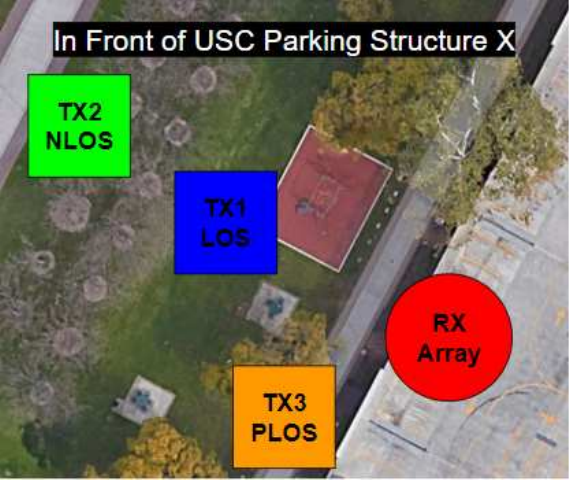}
        \caption{Outdoor}
    \end{subfigure}
\caption{Two different measurement scenarios}%
\label{fig:meas_scene}%
\end{figure}

\subsection{Measurement Scenarios}
For initial verification of the setup functionality, first measurements were conducted in the anechoic chamber right after the calibration. TX antenna and RX array remained in the same position as during calibration (RX array still attached to the positioner), as shown in Fig. \ref{fig:meas_scene} to minimize calibration error. Then, outdoor measurements were conducted, where a TX antenna was 1 m above ground at three different positions and the RX array was on top of a four stories high parking structure (USC PSX building) sitting on top of a cart. The three different positions for TX covered scenarios which were LOS, PLOS (where only parts of the arrays were at LOS due to a waist-high wall - example: only 1st row of all 4 rows was at LOS), and NLOS, where the LOS connection was blocked by trees. Lastly, the sounder was moved back onto the positioner, and another LOS measurement was done in the anechoic chamber. As the outdoor measurement studies were conducted two months after the calibration, the objective of this third experiment was to verify whether calibration has been retained, to narrow down potential error sources in our studies.

\section{Frequency Extrapolation} \label{MSE}
This section reviews the principle of HRPE-based channel extrapolation and defines the figures of merit of this study, MSE and RBG, from measurement data and calibration data. In brief, we extract from measurements within a ``training bandwidth'' the MPC parameters through HRPE, and use them to synthesize the transfer function in the desired frequency band; the figures of merit describe the difference between these extrapolated transfer functions and the ground truth, \textit{i.e.}, the actually measured transfer function in the desired band.  

\begin{figure}[hbt!]
    \centering
    \includegraphics[width=7.9cm]{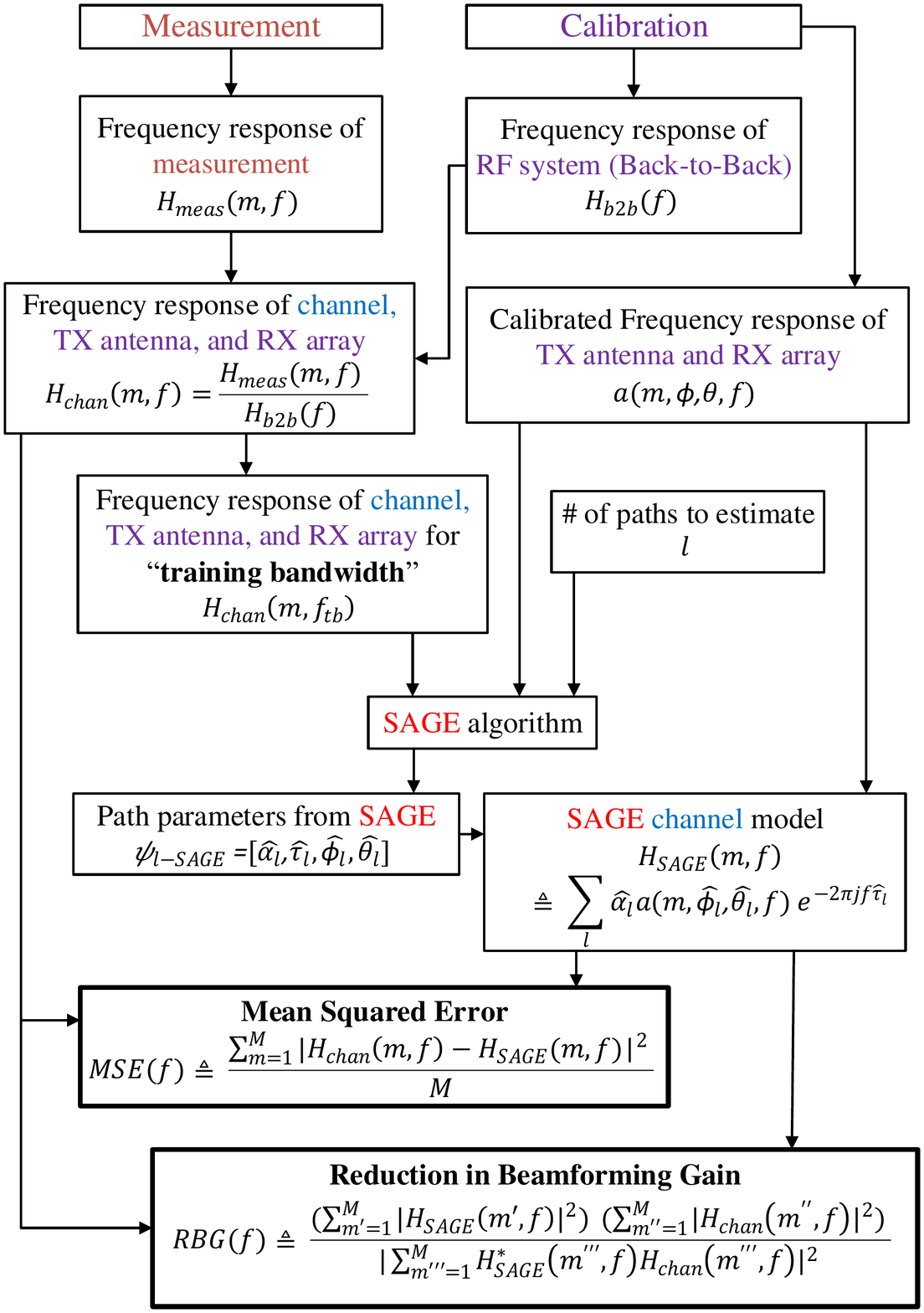}
    \caption{Process of frequency extrapolation}
    \label{fig:proc}
\end{figure}

Figure \ref{fig:proc} outlines the detailed steps. While the calibration data obtained with the VNA provides the frequency response directly, the (time-domain) measurement data received from the ADC are Fourier-transformed (after cyclic prefix removal) to provide the frequency response. It is averaged over the waveform repetitions to improve the SNR, providing $H_{meas}(m,f)$ where $m$ is the port number and $f$ is frequency in Hz (there are in total 64 ports and 2801 frequency points, as specified in TABLE \ref{Sounder}). The frequency response of the RF chain obtained through RF back-to-back calibration (see Section \ref{cali}), $H_{b2b}(f)$, is removed from $H_{meas}(m,f)$ to provide $H_{chan}(m,f)$, which is a combined response of channel, TX antenna, and RX array.

In order to analyze channel extrapolation, a ``training bandwidth'', \textit{i.e.}, a subset of total measurement bandwidth, is selected (in a practical system this would be the bandwidth over which the BS can observe uplink pilots). Among all measured frequency points, those within this training bandwidth are referred as $f_{tb}$. In this particular study, the first 35 MHz (3.325 - 3.360 GHz) of 350 MHz bandwidth was chosen, which corresponded to 281 of 2801 frequency points. The subset of $H_{chan}(m,f)$, denoted as $H_{chan}(m,f_{tb})$, and the calibrated complex pattern of TX antenna and RX array, $a(m,\phi,\theta,f)$ (see Section \ref{cali}), are used as inputs for SAGE algorithm, a widely popular HRPE algorithm \cite{SAGE}.

The SAGE algorithm models the transfer functions at the different antenna elements as the sum of plane waves (MPCs): 
\begin{equation}
    H_{SAGE}(m,f) \overset{\Delta}{=} \sum_{l=1}^{L}\hat{\alpha}_{l} a(m, \hat{\phi}, \hat{\theta}, f)e^{-2\pi j f \hat{\tau_l}}.
    \label{H_SAGE}
\end{equation}
This model has a number of important implicit assumptions: \textit{e.g.}, that the absolute amplitude of an MPC is constant across the different antenna elements, and that no wave-front curvature exists. A further important question is the number of modeled paths $L$. A larger $L$ might be required to represent a sufficient percentage of the total field; yet increasing the number of estimated parameters might also increase the estimation errors due to over-fitting. Thus, while $L \le 4$ might be sufficient for extrapolation if the scenario is very simple (example: LOS scenario in anechoic chamber), complicated scenarios may require as much as $> 20$ paths. Finally, we note that SAGE is an {\em iterative} algorithm that might converge to a local minimum, depending on initialization and various iteration parameters; for more details see, \textit{e.g.}, \cite{SAGE}.

The output from SAGE are the parameters of the MPCs, which include complex amplitude, delay, azimuth and elevation ($\psi_{l-SAGE} = [\hat{\alpha_l}, \hat{\tau_l}, \hat{\phi_l}, \hat{\theta_l}$]). From these parameters, the SIMO channel model is used to reconstruct the channel for each port at each frequency using equation (\ref{H_SAGE}).

Note that this approach provides the extrapolated channel in the {\em uplink}. To be used in an actual system, the channel and associated beamformer needs to be translated to the equivalent {\em downlink} channel. Our measurement setup is not designed to include this step, which is also known as reciprocity calibration. However, it is the same as in TDD systems, where it has been widely explored \cite{Rusek}. Furthermore, any errors occurring in this step would be additive to, and essentially independent of, the errors from the extrapolation procedure.  

Lastly, both $H_{chan}(m,f)$ and $H_{SAGE}(m,f)$ are used to calculate both the MSE and the RBG for the extrapolated channel at any frequency $f$ of interest. First, MSE is calculated as follow:
\begin{equation}
    MSE(f) \overset{\Delta}{=}\frac{\sum_{m=1}^M|H_{chan}(m,f)-H_{SAGE}(m,f)|^2}{M}
\end{equation}
where M is the number of ports evaluated (in our case, $M=64$). If $f$ lies within the training bandwidth, the MSE describes the classical accuracy of the SAGE-extracted MPCs for channel sounding; for $f$ outside the training band, it describes the quality of the channel extrapolation.

Also, the RBG refers to how much loss occurs with knowledge of estimated channel response assuming matched filter. Knowledge of $H_{SAGE}$ is compared to the perfect channel knowledge, which in our study refers to knowledge of $H_{chan}$ (min RBG = 1 = 0 dB). 
\begin{multline}
    RBG(f) \\
    \overset{\Delta}{=}\frac{(\sum_{m'=1}^M|H_{SAGE}(m',f)|^2)(\sum_{m''}^M|H_{chan}(m'',f)|^2)}{|\sum_{m'''=1}^MH_{SAGE}^*(m''',f)H_{chan}(m''',f)|^2}
\end{multline}

\section{Results and Discussions} \label{Results}

\begin{figure}[hbt!]
    \centering
    \begin{subfigure}[b]{0.45\textwidth}
        \centering
        \includegraphics[width=8cm]{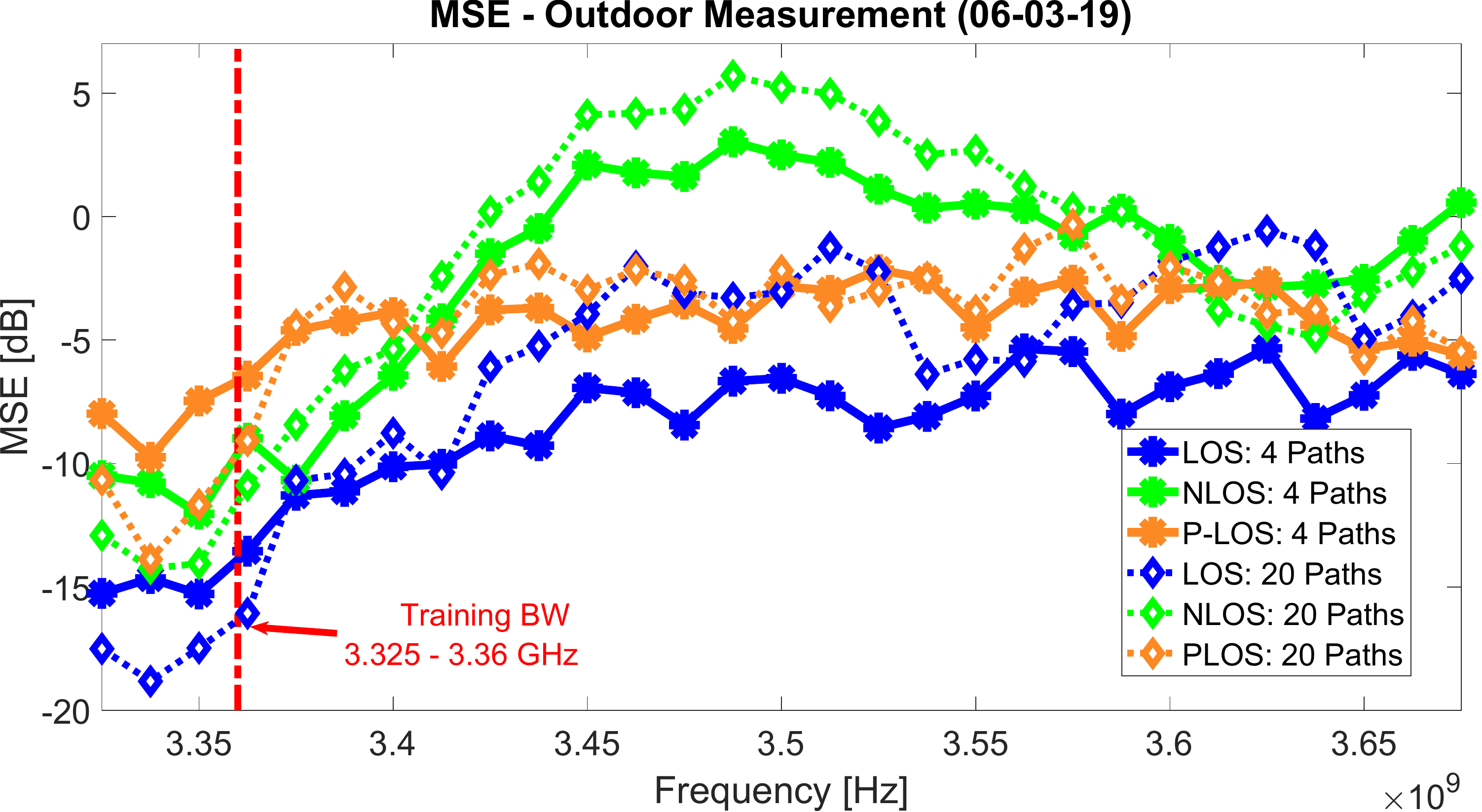}
        \caption{MSE - outdoors}
        \vspace*{2mm}
    \end{subfigure}
    
    \begin{subfigure}[b]{0.45\textwidth}
        \centering
        \includegraphics[ width=8cm]{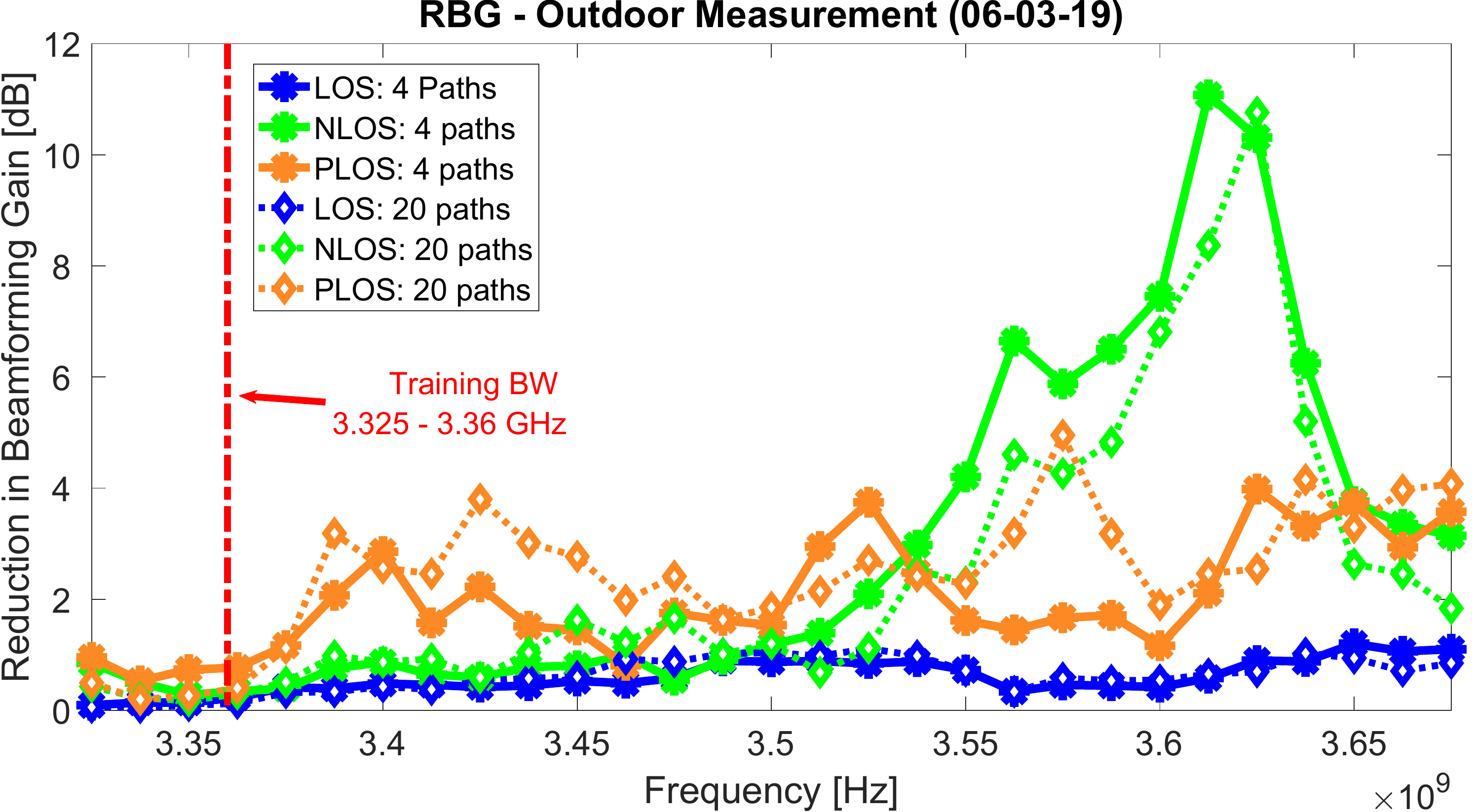} %
        \caption{Reduction in beamforming gain - outdoors}
    \end{subfigure}
\caption{Results from outdoor measurements}%
\label{fig:meas_outdoor}%
\vspace*{7mm}
\end{figure}

\begin{figure}[hbt!]
    \centering
    \begin{subfigure}[b]{0.45\textwidth}
        \centering
        \includegraphics[width=8cm]{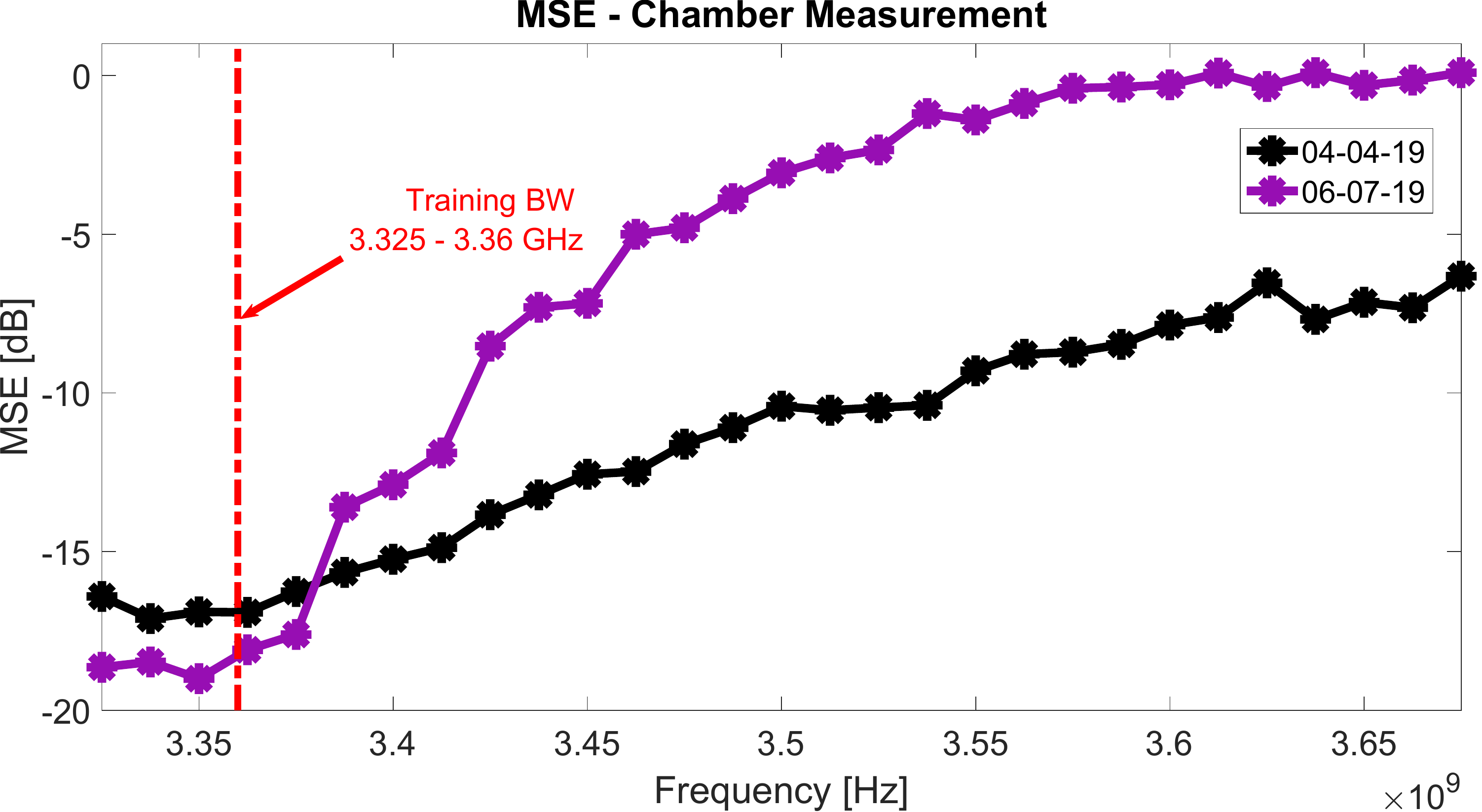}
        \caption{MSE - chamber}
        \vspace*{2mm}
    \end{subfigure}

    \begin{subfigure}[b]{0.45\textwidth}
        \centering
        \includegraphics[ width=8cm]{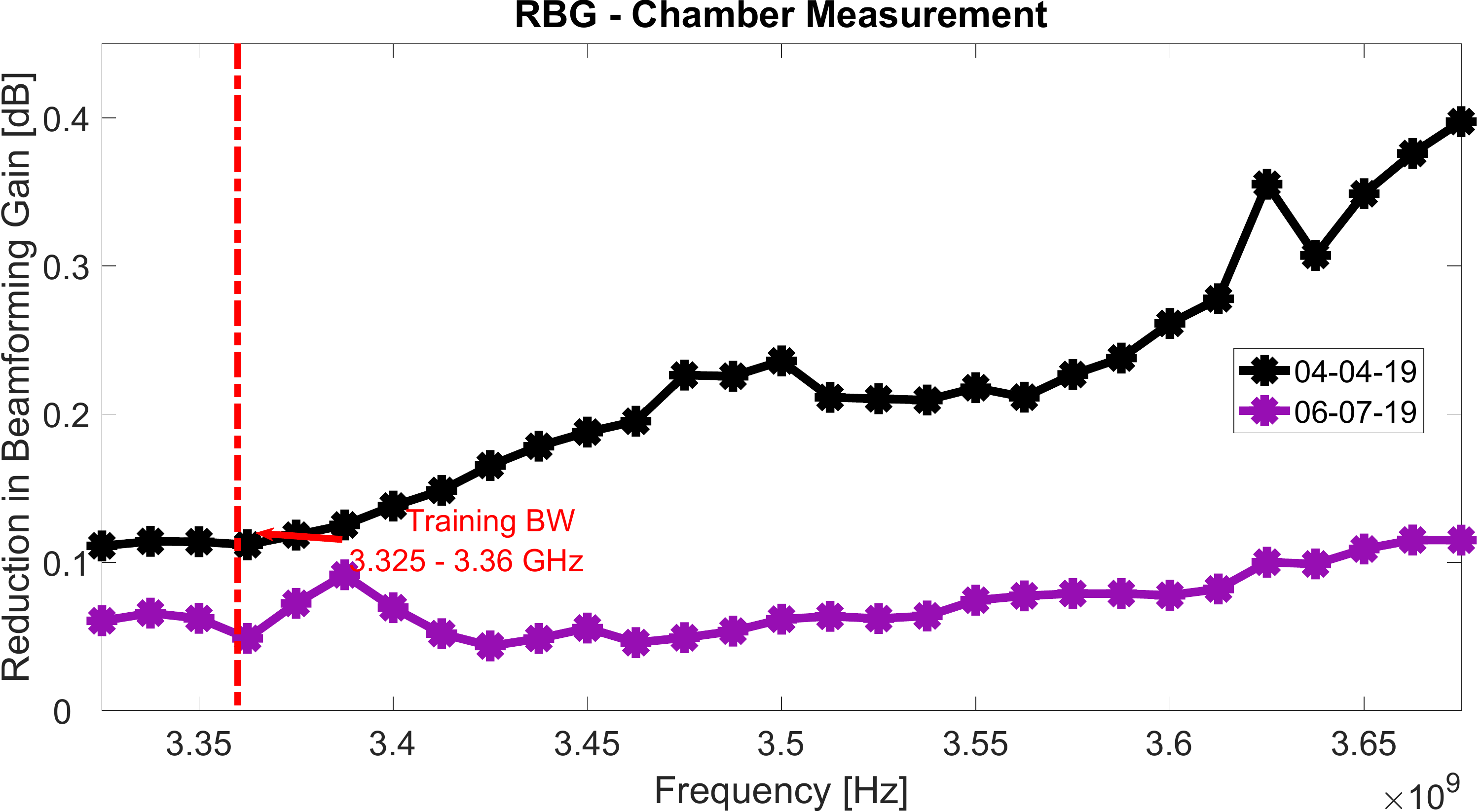} %
        \caption{Reduction of beamforming gain - chamber}
    \end{subfigure}
\caption{Results from chamber measurements (LOS)}%
\label{fig:meas_chamber}%
\end{figure}

Fig. \ref{fig:meas_outdoor} and Fig. \ref{fig:meas_chamber} show results of MSE and RBG in the outdoor and chamber environments (Fig. \ref{fig:meas_scene}). We see that outside the training bandwidth, the MSE is usually larger than $-10$ dB, and often approaches or even exceeds $0$ dB for NLOS case. We discuss in the following some main insights on the channel properties the error depends on.

First, we consider the error {\em within} the training bandwidth ($3.325 - 3.36$ GHz), henceforth called {\em reconstruction error}. Its value depends on the particular scenario: for the LOS cases, MSE was $< -15$ dB, but increases as the channel moves from LOS, to NLOS. This may be due both to a reduction of the SNR, and the fact that a richer multipath environment makes the estimation of the MPC parameters more difficult due to interpath interference. We also see that the reconstruction error is even larger in the PLOS channel, where some rows are at LOS and other rows are at NLOS. This can be explained by the fact that PLOS violates the fundamental SAGE signal model, as the absolute amplitude of the MPCs across the antenna elements is not constant. 

The MSE outside the training bandwidth (henceforth called {\em extrapolation error}) remains, for the LOS case, $<-10$ dB for up to 70 MHz distance from the 35 MHz training band. Furthermore, the extrapolation error in terms of MSE increases as the channel moves from LOS, to NLOS, to PLOS for up to 40 MHz distance from the training band. 

We also find that for all scenarios an increase in the number of estimated paths $L$ leads to a lower reconstruction error - it is intuitive that a more sophisticated model can better represent the channel within the measured bandwidth; it can even correct partially for model mismatch, e.g., by approximating a curved wavefront by a sum of plane waves. This situation changes when we consider the extrapolation error.  There, the MSE becomes {\em worse} as we increase $L$ from 4 to 20. This is due to the fact that estimation errors in the delay are more pronounced for the additional, weaker (lower SNR) MPCs that are estimated in the latter case; delay estimation errors strongly impact the phase at extrapolated frequencies. The increase of the extrapolation error with increasing $L$ is most pronounced in the LOS case, mostly because for NLOS the error is high overall anyway. 

Besides the MSE, we are also interested in the loss of beamforming gain. Fig. \ref{fig:meas_outdoor}b shows the results for the different environments. We see that for the LOS scenario, the loss is less than 1.2 dB over the whole 350 MHz bandwidth, while for NLOS, the loss in  less than 2 dB until 165 MHz outside the band (3.325 - 3.525 GHz). It is worth remembering that the ideal beamforming gain achieved in our system is $18$ dB, so even a $3$ dB loss is comparatively low. 

Still, the relatively high extrapolation error motivated us to investigate possible error sources. The key candidates were (i) model mismatch, and (ii) calibration errors. In terms of model mismatch, we already noted that the PLOS setup results in a violation of the model assumption of absolute amplitude of the MPCs across the antenna elements being constant, so that both reconstruction error  and extrapolation error are high. However, this does not completely explain the extrapolation error for NLOS, since the reconstruction error for this scenario is smaller than in PLOS, but extrapolation error is much larger. 

To test the calibration error, we had performed simple LOS measurements in the anechoic chamber, see Fig. \ref{fig:meas_chamber}. We see that in the measurements conducted right after the calibration (see black line, denoted by 04-04), the extrapolation error is small - less than $-10$ dB up to 200 MHz distance from the training band. However, for a measurement that was conducted after two months since the calibration (see purple plots, 06-07), performance degraded considerably (note that while the measurement was conducted in the same chamber, the measurements are not exactly comparable due to modified setups within the chamber, as indicated also by the fact that the reconstruction error of the later measurements is actually {\em smaller} than that of the earlier measurements). Furthermore, the MPC parameters extracted from the chamber measurements agree with the physical conditions, \textit{i.e.}, directions and delay of the MPC components. RBG was actually less for (06-07) in comparison to (04-04), despite worse performance in MSE, (in all cases the actual RBG is small anyways due to very simple and optimistic scenario).

This indicates that calibration has been lost to a certain degree in those two months, despite the fact that the array/switch combination was not altered consciously. We cannot exclude the possibility that aging of the material of the patches (copper oxidization), small movement of cables due to vibrations during transport, and other irreproducible effects lead to the change in calibration. We stress that while such miniscule calibration change strongly affected the channel extrapolation error, it did not affect the HRPE parameters, so that it would not be noticed in a regular high-resolution channel sounding campaign. 

\section{Conclusion} \label{Conclusion}

We have provided experimental results for frequency extrapolation of the channel transfer function based on HRPE. Our experiments resulted in an MSE that in most outdoor cases was above $-10$ dB outside 35 Mhz training band, implying that the extrapolation of the instantaneous channel response is highly sensitive to calibration errors. On the other hand, in terms of beamforming gain, the results looked more promising. This implies that a large MSE does not necessarily induce a high loss in beamforming power.

In conclusion, our results do not necessarily imply that the channel extrapolation for FDD massive MIMO cannot be used in practice. Other papers \cite{MIT_FDD, Rice_FDD, deeplearning_FDD} showed its possibilities using other figures of merit and methods. 
For our future works, we will study other figures of merit to evaluate the performance of channel extrapolation using various HRPE algorithms. Also, the methods to keep calibration as consistent as possible will be studied. Lastly, the measurements will be extended to full MIMO system to see effects of number of antennas on both TX and RX ends. 

\vspace{0pt}
\noindent{\bf Acknowledgement:} Part of this work was supported by NSF EECS-1731694, a gift from Samsung America, and the Belgian American Educational Foundation (B.A.E.F.). Measurement help from Zihang Cheng is gratefully acknowledged.

\bibliography{mMIMO.bib}
\bibliographystyle{IEEEtran}

\end{document}